\documentclass[aps,prl,twocolumn,groupedaddress,showpacs]{revtex4}
\usepackage{graphicx}
\usepackage{amsmath,amssymb,amstext}

\newcommand{\rydn}{\textsf{n}}
\begin{document}
\title{Rabi oscillations and excitation trapping\\in the coherent excitation of a mesoscopic frozen {R}ydberg gas}


\author{M. Reetz-Lamour}
\author{T. Amthor}
\author{J. Deiglmayr}
\author{M. Weidem\"uller}
 \email[]{weidemueller@physik.uni-freiburg.de}
 \affiliation{Physikalisches Institut der Universit\"at Freiburg,
Hermann-Herder-Str. 3, D-79104 Freiburg, Germany}

\date{\today}

\begin{abstract}
We demonstrate the coherent excitation of a mesoscopic ensemble of
about 100 ultracold atoms to Rydberg states by driving Rabi
oscillations from the atomic ground state. We employ a dedicated
beam shaping and optical pumping scheme to compensate for the
small transition matrix element. We study the excitation in a
weakly interacting regime and in the regime of strong
interactions. When increasing the interaction strength by pair
state resonances we observe an increased excitation rate through
coupling to high angular momentum states. This effect is in
contrast to the proposed and previously observed
interaction-induced suppression of excitation, the so-called
dipole blockade.
\end{abstract}

\maketitle

Rydberg atoms, with their rich internal structure, have been in
the focus of atomic physics for more than a
century~\cite{gallagher94}. In the last decade Rydberg physics
were extended to laser-cooled atomic gases which allowed the study
of a frozen system with controllable, strong interactions and
negligible thermal contributions. This has opened a wide field in
both experiment and theory covering such diverse areas as resonant
energy transfer~\cite{anderson98,mourachko98}, plasma
formation~\cite{robinson00,amthor07}, exotic
molecules~\cite{boisseau02,greene00}, and quantum random
walks~\cite{cote06,muelken07}. In addition, these frozen Rydberg
systems have been proposed as a possible candidate for quantum
information processing~\cite{jaksch00,lukin01}. However, the
coherent excitation of Rydberg atoms, an important prerequisite
for quantum information protocols, has proven a challenging task.
While Rabi oscillations between different Rydberg states have been
demonstrated and thoroughly analyzed before~\cite{gentile89}, Rabi
oscillations between the ground state and Rydberg states of atoms
have not been observed directly so far, mainly owing to the small
transition matrix element.

In this Letter we report on the experimental realization and
observation of Rabi oscillations between the ground and Rydberg
states of ultracold atoms. We demonstrate how strong interatomic
interactions influence the coherent excitation of a mesoscopic
cloud of atoms. We show that an interaction-induced coupling to a
larger number of internal states can be used to trap the
excitation. Employed in a controlled way, this effect offers
future applications in the experimental realization of quantum
random walks with exciton trapping~\cite{muelken07}.

\begin{figure}
\includegraphics[width=\columnwidth]{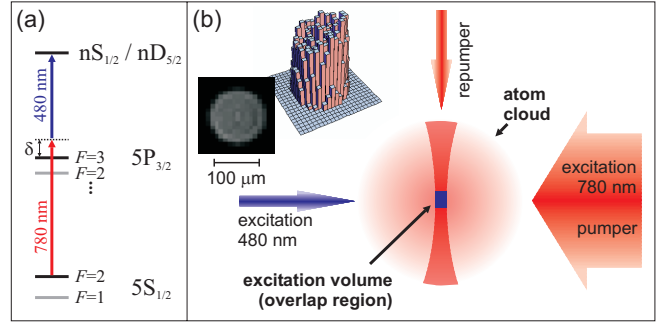}
\caption{(a) Relevant level scheme for two-photon excitation of
rubidium. (b) Preparation of a mesoscopic subensemble. All atoms
are pumped to the $F$=1 hyperfine component of the ground state.
Only a tight tube of atoms is pumped to $F$=2 wherefrom the
excitation originates. A mesoscopic subensemble containing about
100 atoms within this tube is transferred to a Rydberg state by
near-resonant two-photon excitation with two counterpropagating
laser beams at 780\,nm and 480\,nm, respectively. The excitation
laser at 480\,nm has a flat-top intensity profile shown in the
inset. This ensures a constant Rabi frequency over the excitation
volume.\label{fig:setup}}
\end{figure}

Our experiments are performed with a magneto-optical trap (MOT) of
about $10^7$ $^{87}$Rb atoms at densities of
$10^{10}\,\mathrm{cm}^{-3}$ and temperatures below 100\,$\mu$K.
Rydberg excitation is achieved with two counterpropagating laser
beams at 780\,nm and 480\,nm (see Fig.~\ref{fig:setup}). The laser
at 780\,nm is collimated to a waist of 1.1\,mm ensuring a constant
Rabi frequency of $2\pi\times 55\,$MHz over the excitation volume
as determined from Autler-Townes splittings~\cite{deiglmayr06}.
The laser at 480\,nm is referenced to a temperature stabilized
\textsc{Zerodur}-resonator and its beam is shaped with a
diffractive optical element that produces a flattop beam profile
which is characterized with an adapted CCD camera with a spatial
resolution of 5.6\,$\mu$m. The measured flattop is depicted in
Fig.~\ref{fig:setup}. Further details will be described
elsewhere~\cite{reetz07}. This setup allows a tightly focussed
beam with high intensity and at the same time overcomes the limits
of a Gaussian beam shape. Gaussian beams have been used for
stimulated adiabatic passage to high Rydberg
states~\cite{cubel05,deiglmayr06} but do not allow the observation
of synchronous Rabi floppings in a mesoscopic sample due to the
wide range of intensities across the beam~\cite{deiglmayr06}. In
order to address only atoms in the flattop region we use spatially
selective optical pumping (see Fig.~\ref{fig:setup}): The magnetic
field of the MOT is turned off 3\,ms before excitation
\footnote{We use the thermal expansion of the cloud to change the
ground state density by increasing this time up to 15\,ms.}. After
turning off the trapping lasers, all atoms are pumped to the $F$=1
hyperfine component of the 5S$_{1/2}$ ground state within
300\,$\mu$s. A small tube of atoms perpendicular to the excitation
beams is transferred to the $F$=2 hyperfine component with a
1\,$\mu$s pulse of a tightly focused repumping beam (waist
10\,$\mu$m, 10\,nW). Only these atoms are resonant with the
excitation lasers. They are optically pumped to the stretched
$F$=2, $m_F$=2 state by a 1$\mu$s pulse of a $\sigma^+$-polarized
pumper beam superimposed with the excitation lasers. From this
state we can excite both stretched \rydn S and \rydn D states
depending on the helicity of the excitation laser polarization;
\rydn~denotes the principal quantum number. By maintaining the
two-photon resonance but detuning from the intermediate state
5P$_{3/2}$ by typically $\delta\,/\,2\pi=140\,$MHz this state
experiences negligible population and atoms are directly
transferred to the Rydberg state~\cite{deiglmayr06}. The overlap
between the repumping laser and the 480\,nm laser defines the
excitation volume of about $10\,\mu\mathrm{m} \times
10\,\mu\mathrm{m} \times 100\,\mu\mathrm{m}$, corresponding to
$\sim$100 atoms (see Fig.~\ref{fig:setup}.{\footnotesize (b)}).
Rydberg atoms are detected by field-ionization and acceleration
onto a microchannel plate detector. There is no signal for
electric fields below the ionization threshold, therefore we can
rule out the presence of spurious ions during excitation. After
detection the MOT is turned back on and the whole cycle is
repeated every 70\,ms. Our setup allows to compensate electric
fields $\mathcal{E}_\perp$ in a direction perpendicular to the
excitation lasers. Field components parallel to the field plates
cannot be compensated and are determined from Stark shifts to be
$\mathcal{E}_\parallel$=160\,mV/cm~\cite{singer05a}. The field
values given in the text represent the total field
$\mathcal{E}=(\mathcal{E}_\parallel^2+\mathcal{E}_\perp^2)^{1/2}$.

\begin{figure}
\includegraphics[width=\columnwidth]{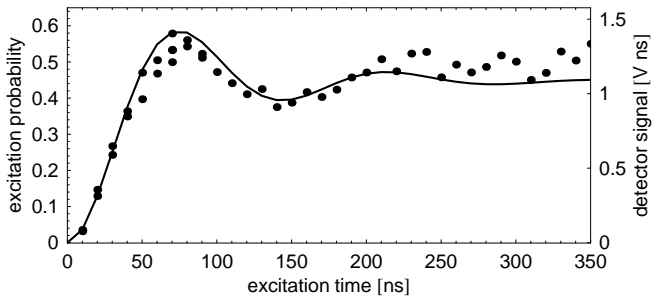}
\caption{Rabi oscillations between the 5S$_{1/2}$ and 31D$_{5/2}$
state of $^{87}$Rb. Each dot is an average of measurements over 28
experimental repetition cycles. The solid line shows the simulated
excitation probability taking the measured intensity distribution,
the residual admixture of the intermediate 5P$_{3/2}$ state, and
our finite laser linewidth into account. The scaling between
detector signal and simulated excitation probability is a free
parameter that represents the detector
efficiency.\label{fig:Rabi}}
\end{figure}

Fig.~\ref{fig:Rabi} shows the measured fraction of excited Rydberg
atoms in the 31D$_{5/2}$ state as a function of excitation time.
Rabi oscillations are clearly visible. We observe the same
temporal dependance for both \rydn S and \rydn D states with
\rydn$<$40 at both low and high atom densities. This indicates
that for these small \rydn~values the observed Rabi oscillations
are not affected by Rydberg--Rydberg interactions and that all
atoms perform identical Rabi oscillations simultaneously. The
solid line in Fig.~\ref{fig:Rabi} is a theoretical prediction
which takes the measured time dependance of the excitation pulses
into account. It incorporates three sources of dephasing: the
remaining admixture of the intermediate state, the residual
intensity distribution of the flat-top beam profile, and a finite
laser bandwidth of 2.4\,MHz. With increasing importance these
three sources lead to the observed damping of the measured Rabi
oscillations. A detailed discussion together with quantitative
comparisons between measured and calculated Rabi frequencies and
other systematic verifications will be published
elsewhere~\cite{reetz07}. For longer times ($\gtrsim 200\,$ns) the
excitation probability reaches a steady state value of 1/2. This
allows us to calibrate the detector: by overlapping the model with
the measured data we obtain a scaling factor between single atom
excitation probability and detector signal \footnote{This factor
is inverse proportional to the number of atoms in the excitation
volume and therefore also to the density. By determining this
factor we can thus calibrate relative densities. We do this for
low-\rydn\ states. Absolute densities are determined by absorption
imaging.}.

The observation of Rabi oscillations between ground and Rydberg
states is a first step towards the implementation of quantum
information protocols with cold Rydberg gases. The most promising
protocols further rely on the so-called dipole
blockade~\cite{lukin01}. The dipole-blockade describes a system in
which the excitation of more than one atom is detuned from the
single-atom excitation due to interaction-induced energy shifts,
which suppresses multiple excitations. It allows to store a single
excitation in a mesoscopic cloud of atoms and is accompanied by an
increased collective Rabi frequency, evidence of which was
observed only recently~\cite{heidemann07}. The first experimental
signatures of the blockade were observed as a suppression of
excitation~\cite{tong04,singer04} due to the extraordinarily
strong van der Waals interactions at large values \rydn. Recently,
a similar effect was observed at lower values of \rydn~by using
pair state resonances to switch from van der Waals to the stronger
resonant dipole-dipole interaction with an external electric
field~\cite{vogt06}.

\begin{figure}
\includegraphics[width=\columnwidth]{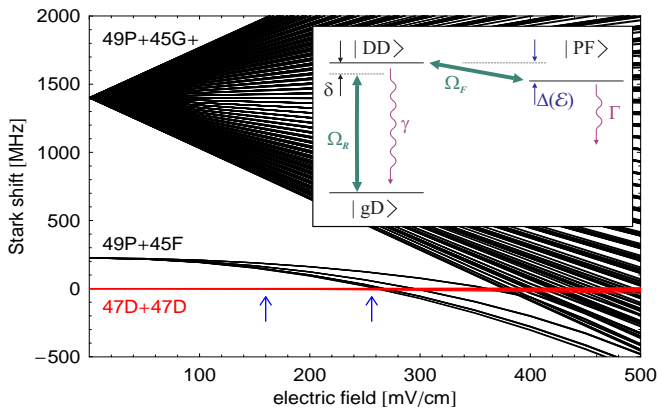}
\caption{Stark shift of dipole coupled pair states around the
47D$_{5/2}$+47D$_{5/2}$ asymptote. The two arrows mark the total
electric field at which the two spectra of Fig.~\ref{fig:Spectra}
are taken. At an electric field of $\sim260\,$mV/cm the
47D$_{5/2}$+47D$_{5/2}$ state is degenerate with the
49P$_{3/2}$+45F$_J$ ($J$=5/2, 7/2) state which leads to resonant
dipole-dipole interaction. The inset shows a simplified level
scheme for this system involving the coherent excitation of a
ground state/Rydberg state atom pair $\left|\mathrm{gD}\right>$ to
a pair of two Rydberg atoms $\left|\mathrm{DD}\right>$ with a Rabi
frequency $\Omega_R$, a laser detuning $\delta$ and an effective
linewidth $\gamma$. The state $\left|\mathrm{DD}\right>$ is dipole
coupled to an almost degenerate pair $\left|\mathrm{PF}\right>$
with an interaction energy $\hbar\,\Omega_F$ and an energy
difference $\hbar\,\Delta$ that depends on the electric field
$\mathcal{E}$. The dephasing of the dipole coupling (see text) is
phenomenologically described by a decay rate $\Gamma$ out of the
three-level system.\label{fig:StarkMap}}
\end{figure}

\begin{figure}
\includegraphics[width=\columnwidth]{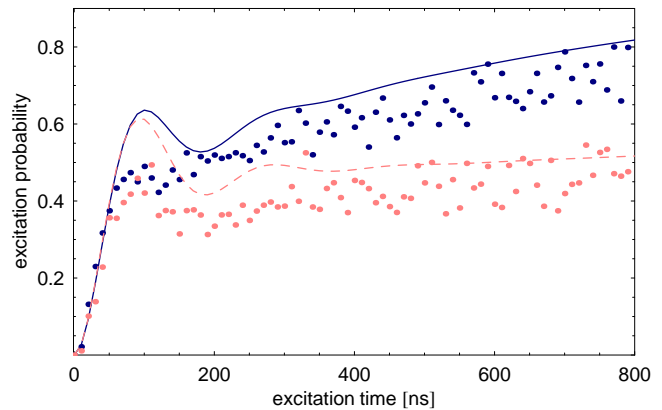}
\caption{Temporal evolution of the excitation probability for the
47D$_{5/2}$ state at an electric field $\mathcal{E}=160$\,mV. For
the smaller density of $4.7\times 10^{9}\,\mathrm{cm}^{-3}$ (gray
dots) we see a damped Rabi oscillation similar to
Fig.~\ref{fig:Rabi}. For the higher density of
$10^{10}\,\mathrm{cm}^{-3}$ (black dots) the stimulated emission
is suppressed leading to a constantly increasing Rydberg
population. The lines are the solution of the model schematically
shown in Fig.~\ref{fig:StarkMap} for weak (dashed) and strong
(solid) interactions.\label{fig:Time}}
\end{figure}

For rubidium, \rydn D states offer similar resonances.
Fig.~\ref{fig:StarkMap} shows the electric field dependance of the
involved pair states. At vanishing electric field two 47D$_{5/2}$
atoms experience van der Waals interaction. At an electric field
of $\sim260\,$mV/cm the 47D$_{5/2}$--47D$_{5/2}$ pair state
becomes energetically degenerate with the dipole-coupled
49P$_{3/2}$--45F$_J$ ($J$=5/2, 7/2) pair state. This changes the
interaction to long-range resonant dipole interaction. In contrast
to the picture of the dipole blockade, however, the increased
interaction strength does not translate into a reduced excitation
probability at the atomic resonance: Fig.~\ref{fig:Time} shows the
temporal evolution of the 47D$_{5/2}$ population for two different
densities. For the lower density of $4.7\times
10^{9}\,\mathrm{cm}^{-3}$ (gray points) we observe damped Rabi
floppings similar to low-\rydn~states. For the higher density
($10^{10}\,\mathrm{cm}^{-3}$, black points) we observe an
increased excitation that clearly exceeds 0.5. This means that
part of the atoms remain in Rydberg states without being
stimulated back to the ground state. The effect is clearly
density-dependent and must therefore be interaction-induced.

As our detection cannot distinguish between 47D, 49P, and 45F, we
assume an accumulation of atoms in the latter states. However, the
increased excitation is in contrast to the previously observed
excitation blockade at another pair state resonance~\cite{vogt06}.
We attribute the difference to the fact that the pair state
resonance used in Ref.~\cite{vogt06} comprises well-defined pair
states (\rydn P$_{3/2}$--\rydn P$_{3/2}$ and \rydn
S$_{1/2}$--(\rydn+1)S$_{1/2}$ with $\left| m \right|=1/2$) while
in our case the involved 45F$_J$
state is highly degenerate. As each sub-state has a different
coupling strength this leads to a dephasing and a coupling back to
the original 47D state is suppressed. The residual electric field
in our setup enhances this dephasing as it mixes different
$m_J$-levels and thus leads to a dipole coupling with all possible
values of $m_J$, comprising 14 sub-states of the two values of
$J=5/2$ and $7/2$ for the 45F$_J$ state and 6 sub-states for the
47D$_{5/2}$ state.

We can heuristically model the dephasing between different pair
states with an analytical model illustrated in
Fig.~\ref{fig:StarkMap}. It incorporates the coherent excitation
of a pair $\left|\mathrm{gD}\right>$ of a ground state atom
$\left|\mathrm{g}\right>$ and a 47D atom $\left|\mathrm{D}\right>$
to a pair $\left|\mathrm{DD}\right>$ of two 47D atoms as well as
the coupling to the $\left|\mathrm{PF}\right>$ state with a
field-dependent energy shift relative to
$\left|\mathrm{DD}\right>$. The dephasing is introduced
phenomenologically by a decay rate $\Gamma$ out of the three level
system. This system is an analog of the excitation of an optical
three level system and can be fully described by the according
optical Bloch equations~\cite{whitley76}. In the case of coherent
couplings ($\Gamma=0$), this system describes the dipole blockade
as an analog to the Autler-Townes splitting in the optical system.
By including incoherent coupling ($\Gamma>0$), we reproduce the
increased excitation measured in Fig.~\ref{fig:Time}: The solid
line corresponds to model calculations with parameters for the
excitation that are scaled from measured low-\rydn~Rabi
oscillations. We have set $\Delta$ to the theoretical value
$\Delta(\mathcal{E}=160\,\mathrm{mV})=-2\pi\times150$\,MHz (see
Fig.~\ref{fig:StarkMap}). This leaves two free parameters: the
coupling strength $\Omega_F=2\pi\times 40$\,MHz and the dephasing
rate $\Gamma=2\pi\times 30$\,MHz where chosen to achieve a good
overlap with the experimental data. Both values are consistent
with the theoretical value $\Omega_{F}^{max}=2\pi\times 64$\,MHz
for the strongest coupling between the stretched states
($m_J=J=\ell+1/2$) at the most probable interatomic distance of
2.6\,$\mu$m. The dashed line corresponds to a low density model
where $\Omega_F$ and $\Gamma$ are scaled down proportional to the
density (as expected for an $1/R^3$-dependent interaction).

\begin{figure}
\includegraphics[width=\columnwidth]{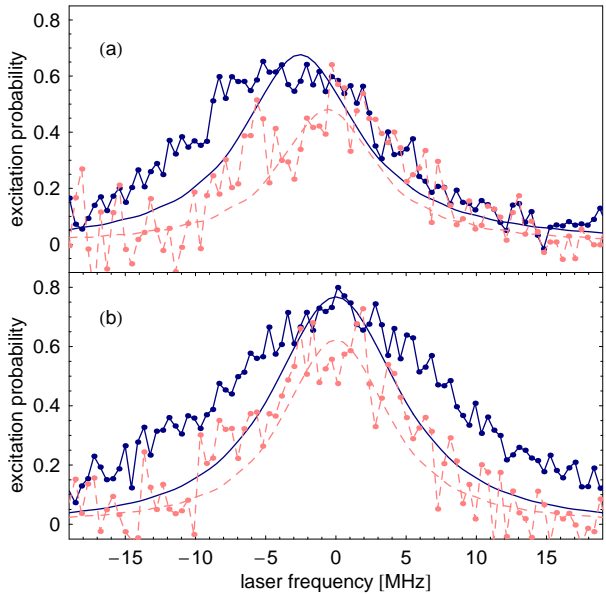}
\caption{(a) Spectrum of the 47D$_{5/2}$ resonance line at an
electric field of $\mathcal{E}$=160\,mV/cm at densities of
$10^{10}\,\mathrm{cm}^{-3}$ (blue) and $4.7\times
10^{9}\,\mathrm{cm}^{-3}$ (red). For the higher density we observe
a line shift and broadening combined with an increased excitation
probability. The solid and dashed lines are model calculations.
The line shift is a result of an energy difference between the
$\left|\mathrm{DD}\right>$ and $\left|\mathrm{PF}\right>$ pair
states. (b) Same for $\mathcal{E}$=260\,mV/cm. The line shift is
reduced to zero, while line broadening and an increased excitation
remain. All spectra are taken by scanning the 480\,nm laser (fixed
detuning from the intermediate level) with an excitation time of
400\,ns.\label{fig:Spectra}}
\end{figure}

With the same values for $\Omega_F$ and $\Gamma$ our model also
explains spectral measurements shown in Fig.~\ref{fig:Spectra}. By
changing the electric field we can tune the energy difference
between the $\left|\mathrm{DD}\right>$ and
$\left|\mathrm{PF}\right>$ pair states. Fig.~\ref{fig:Spectra}(a)
shows a measured spectrum of the 47D$_{5/2}$ line at
$\mathcal{E}$=160\,mV/cm ($\mathcal{E}_\perp$=0) together with
model calculations. All model parameters are the same as those for
the excitation time scan in Fig.~\ref{fig:Time}.
Fig.~\ref{fig:Spectra}(b) shows a spectrum at
$\mathcal{E}$=260\,mV/cm corresponding to $\Delta$=0. The dashed
line shows the model result in the low-density limit
($\Omega_F=0$), the solid line includes a coupling with
$\Omega_F=2\pi\times 5$\,MHz and $\Gamma=2\pi\times 30$\,MHz, both
chosen to yield the best overlap with the experiment. We attribute
the reduced coupling strength $\Omega_F$ to the increased electric
field which splits the involved states, effectively reducing
$\Omega_F$ from 40\,MHz at 160\,mV to 5\,MHz at 260\,mV. There are
three distinct features of the measured spectra at higher
densities: line broadening, an enhanced excitation probability,
and an electric field dependent shift. The model qualitatively
describes all of these features.

We want to emphasize that the excitation trapping described here
is not due to state redistribution by collisions with
ions~\cite{dutta01} since we have verified that there are no ions
produced during the short excitation time. Furthermore the
excitation volume is too small to support avalanche ionization.
The excitation enhancement is also different from the antiblockade
for the case where the Autler-Townes splitting of the lower
transition matches the Rydberg--Rydberg interaction
energy~\cite{ates07}. The mechanism presented here can be seen as
an energy diffusion \emph{inside} the atom rather than a spatial
diffusion which relies on internal state swapping \emph{between}
atoms and is the source of a different dephasing mechanism for
energy transfer processes~\cite{anderson98,mourachko98}. The
internal energy dissipation can even counteract spatial diffusion
by transferring the excitation into states with different angular
momenta. This can result in pair states which are not
\emph{dipole}-coupled to other pair states thus stopping spatial
diffusion. This effect can be exploited as an exciton trap for
continuous time quantum random walk experiments~\cite{muelken07}.
It also constitutes another constraint to states that are suited
for dipole blockade experiments besides the already identified
zeroes in Rydberg--Rydberg interactions for certain internal
states~\cite{walker05} and atom pair alignments~\cite{reinhard07}.

\begin{acknowledgments}
We thank T.~Pohl for fruitful discussions and C.~Pruss and
W.~Osten from the Institute for Technical Optics in Stuttgart for
the production of the beam shaping element. The project is
supported by the Landes\-stiftung Baden-W\"urttemberg within the
''Quantum Information Processing'' program, and by a grant from
the Ministry of Science, Research and Arts of Baden-W\"urttemberg
(Az: 24-7532.23-11-11/1).
\end{acknowledgments}

After submission of this article we became aware of a
complementary work which demonstrates Rabi oscillation with an
interaction-dependent damping in microscopic samples of 1 to 10
atoms \cite{johnson07}.



\end{document}